# Unidirectional, dual-comb lasing under multiple pulse formation mechanisms in a passively mode-locked fiber ring laser

**Ya Liu,**[1,2] **Xin Zhao,**[1] **Guoqing Hu,**[1] **Cui Li,**[1] **Bofeng Zhao,**[1] **and Zheng Zheng**[1,2,*]

[1] *School of Electronic and Information Engineering, Beihang University, 37 Xueyuan Rd, Beijing 100083, China*
[2] *Collaborative Innovation Center of Geospatial Technology, 129 Luoyu Road, Wuhan 430079, China*
*\*Corresponding author: zhengzheng@buaa.edu.cn*

**Abstract:**

Dual-comb lasers from which asynchronous ultrashort pulses can be simultaneously generated have recently become an interesting research subject. They could be an intriguing alternative to the current dual-laser optical-frequency-comb source with highly sophisticated electronic control systems. If generated through a common light path traveled by all pulses, the common-mode noises between the spectral lines of different pulse trains could be significantly reduced. Therefore, coherent dual-comb generation from a completely common-path, unidirectional lasing cavity would be an interesting territory to explore. In this paper, we demonstrate such a dual-comb lasing scheme based on a nanomaterial saturable absorber with additional pulse narrowing and broadening mechanisms concurrently introduced into a mode-locked fiber laser. The interactions between multiple soliton formation mechanisms result in unusual bifurcation into two-pulse states with quite different characteristics. Simultaneous oscillation of pulses with four-fold difference in pulsewidths and tens of Hz repetition rate difference is observed. The coherence between these spectral-overlapped, picosecond and femtosecond pulses is further verified by the corresponding asynchronous cross-sampling and dual-comb spectroscopy measurements.

## 1. Introduction

Optical solitons not only are fascinating physical phenomena but also possess important practical applications. Mode-locked lasers have been one of most important platforms to explore various soliton effects. As this kind of solitary wave is generated from, in essence, a non-conservative system involving constant exchanging of energy, such lasers' output is often characterized as dissipative solitons (DSs). Unlike optical solitons in the conservative systems under equilibrium, their characteristics and dynamics can be remarkably different, depending on the balance between not only dispersion and Kerr effect but also other gain/loss and linear/nonlinear effects [1, 2]. Therefore, soliton formation in these lasers could be significantly affected by the laser parameters and conditions.

For passively mode-locked lasers, the mode-locking scheme has a profound effect on the pulse characteristics generated, such as the spectral and pulse width. The traditional nonlinear polarization rotation (NPR) scheme relies on the intensity-dependent rotation of the polarization state in the optical fiber, which requires sufficient amount of nonlinearity. Such lasers often have relatively high threshold levels. Nonlinear mode-lockers based on saturable absorption (SA) effect are another alternative to realize low-threshold, self-starting mode-locking. The emergence of carbon nanotube, graphene and, more recently, topological insulator materials bring in a new level of cost-effectiveness and fabrication simplicity [3-5]. Yet, the intensity-dependent transmission response of the material often limits the achieveable pulsewidth. Hybrid schemes had been proposed to achieve higher power outputs and shorter pulses by combining Kerr mode-locking effect and a SA mode-locker [6, 7]. By introducing an additional pulse shorting mechanism often using a polarizer or an absorber with polarization selectivity, significantly shorter pulses than those by the SA scheme alone could be generated, while keeping many advantages of the latter.

Traditionally, most mode-locked lasers are operated and modelled under the condition where only one pulse is emitted during one roundtrip time of the cavity or multiple *identical* pulses under an increased energy in the laser cavity [8, 9]. While the spectrum of the pulse would slightly vary under different pump powers, by further ramping up the pump, a bifurcation to a double-pulse stationary output [10] would occur, which consists of two identical pulses with a slightly narrower bandwidth. This kind of multi-pulsing dynamics has been observed in a variety of laser configurations based on either NPR or SA mode-lockers.

Now, there are increasing interests in exploring the simultaneously emission of multiple *diverse* pulses from one laser cavity. Recently, it had been demonstrated that soliton fiber lasers can generate two or more ultrashort pulses circulating in the same direction in the cavity with different center wavelengths or states of polarization [11, 12]. While traveling along the same physical path, chromatic or polarization-mode dispersion results in different roundtrip time from one cavity. Multi-pulse lasing had also been observed in bidirectionally oscillating cavities [13, 14] or a unidirectional one with slightly different beam paths [15].

In this paper, we demonstrate the simultaneous generation of multiple picosecond and femtosecond pulses with four-fold different pulsewidths from a unidirectional ring laser under the synergistic interactions between multiple pulse formation mechanisms. These pulses have the same polarization state and completely overlapped spectrum. Their small stable repetition rate difference and good coherence render them a possible alternative dual-comb source.

## 2. Experimental setup

The laser cavity setup for unidirectional dual-comb mode-locking is shown in Fig. 1. The total length of the cavity is ~9.0 m, which consists of a 2.1-meter-long piece of Erbium-doped fiber (EDF, Type 1022) forward pumped by a 980 nm pump diode, an inline polarizer aligned roughly along the fast axis of two 1-meter-long polarization maintaining fiber (PMF) pigtails, as well as other single mode fiber (SMF) pigtails of components. A polarization-independent isolator (ISO) and a fiber-squeezer-type polarization controller (PC) are used to ensure the unidirectional operation and control the intracavity polarization evolution, respectively. An 80/20 output coupler (OC) is used. A single-wall carbon nanotube (SWNT) modelocker is fabricated on an FC/APC ferrule using the optical deposition method [16, 17] with an insertion loss of about 1.5 dB at 1550 nm. With its relatively fast picosecond nonlinear saturable absorption [18], SWNT helps to initiate the mode-locking process. The total GVD is ~0.11 ps/nm, and the laser could operate in the soliton regime, where the chirp and spectral broadening by the 3rd-order fiber nonlinearity is offset by that from the anomalous dispersion.

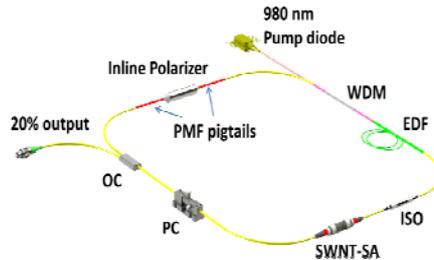

Fig. 1. Configuration of the unidirectional dual-comb mode-locked fiber laser.

Besides these well-known effects in soliton fiber lasers, the introduction of the in-line polarizer with birefringent PMF pigtails in this seemingly simple design bring into plays additional mode-locking and pulse formation mechanisms [19]. In contrast to the traditional hybrid mode-locking lasers based on a mode-locker and the NPR effect, instead of waveplates and a polarizer, the PMF of the inline polarizer introduces relatively large birefringence into the cavity that results in an *ostensibly small yet critical* difference in the pulse formation process. When combined with the polarization-dependent loss from the polarizer, spectral filtering effect would emerge for the imperfectly aligned fiber polarizer. Based on the fiber parameters, its free spectral range (FSR) is about 5 nm at 1560 nm. While NPR is to further narrow the pulses as an ultrafast 'modelocker', filtering would curtail the spectral bandwidth of pulses and, thus, may broaden the pulses. Since the shape of a mode-locked laser output is affected by the delicate balance between the pulse formation mechanisms in the cavity, the simultaneous existence of multiple intracavity linear or nonlinear, pulse narrowing and broadening mechanisms could enable emission of different pulses from the same platform.

### 3. Experimental results and discussions

**Mode-locking under the traditional soliton regime** When PC is adjusted to a 'zero' position so that the state of polarization at the polarizer remains the same after one round trip in the cavity, the polarization-related filtering and NPR pulse narrowing effect as well as the intracavity loss would be minimized. Mode-locking can self-start at a fundamental repetition rate $f_{rep}$ of ~23 MHz. In this case, the optical spectrum and RF spectrum of the fundamental soliton pulses are measured and shown in Fig. 2. They show a typical soliton shape with a 3-dB spectral bandwidth of 2.6 nm, which is mostly determined by the intracavity dispersion, fiber nonlinearity and the mode-locker characteristics. The result is compared with that obtained using a 2-m-long piece of SMF in the place of the inline polarizer, which bears strong similarities. This further suggests that the polarization-related mechanisms are not affecting the pulse forming process in this case.

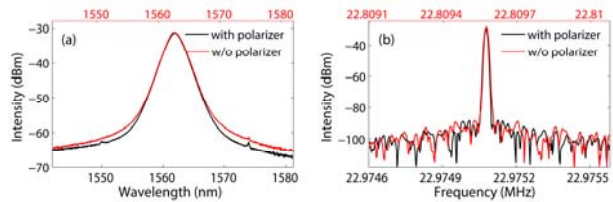

Fig. 2. (a) Optical spectra and (b) RF spectra of the laser under the traditional soliton regime with and without a polarizer.

**Mode-locking under the additional NPR and filtering effects** By adjusting the polarization controller away from the 'zero' position, spectral filtering in the lasing spectrum is observed. Periodic spectral interferometric modulation whose period is determined by the amount of birefringence could limit the pulse bandwidth. Therefore, when the pump power is relatively low, the bandwidth of pulses with relative low energy is reduced from the above level. On the other hand, for pulses with higher pulse energy when the laser is pumped harder, the NPR effect may kick in and, under correct polarization 'bias' positions, it reduces the linear polarization-dependent loss for pulses with higher optical intensity. This pulse narrowing mechanism could balance and eventually overcome the filtering effect. Therefore, even shorter pulses with a broader spectrum can be formed.

As shown in Fig. 3(a), by adjusting PC, a gain peak at ~1562.8 nm can be observed. The laser starts to emit CW light at that wavelength, when the pump power reaches the lasing threshold. Once the pump power further reaches 14.5 mW, a pulse with a spectrum bandwidth of 1.8 nm self-starts to mode-lock at the fundamental cavity frequency. As discussed above, this pulse's narrower spectrum is attributed to the spectral filtering in the absence of the additional hybrid mode-locking effect.

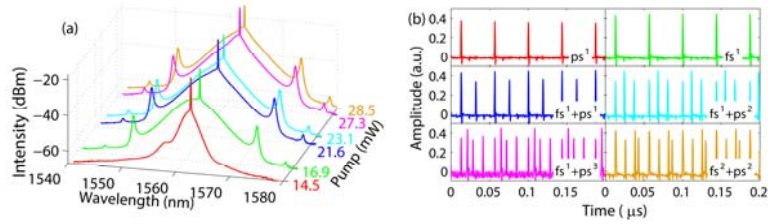

Fig. 3. (a) Optical spectra evolution and (b) the corresponding temporal oscilloscope waveforms, when increasing the pump power.

At a larger pump power while keeping the PC position unchanged, the pulse would accumulate additional energy so that the nonlinear phase shifts it experiences also increase. The pulse's bandwidth broadens to 4 nm at a pump power of 16.9 mW. Its bandwidth indicates the hybrid mode-locking effect in the presence of NPR, as it is significantly wider than that under the soliton regime alone. Higher height of the pulse train when observed on the oscilloscope shown in the top row of Fig. 3(b) also indicates the increased pulse energy. The center wavelength of this shorter pulse shifts slightly to the shorter wavelength and is 0.8 nm away from the CW peak, which is almost unchanged. Significant sidebands are observed, indicating the presence of relatively strong, quasi-CW dispersive wave in the cavity [20, 21].

If further increasing the pump power, without any adjustment to the cavity, a more intriguing phenomenon would occur. An extra 'bump' shows up on top of the optical spectrum of the shorter pulse. Its position is close to the CW peak, and its shape bears a strong resemblance to that of the wider pulse first appeared. The pulse trace shows an additional pulse (denoted as 'ps') besides the taller short pulse (denoted as 'fs'), while the relative temporal position between the 'fs' and 'ps' pulses slightly shifts after each round trip. This indicates that a bifurcation from a single-pulse state to a dual-pulse state has occurred with quite different spectral characteristics, and their repetition rates are also slightly different.

As the pump power is further increased, more pulses are observed in each period of the round trip time. As observed in Fig. 3(b), up to three smaller pulses could co-exist with a taller one, before additional taller pulse would emerge while one smaller pulse 'disappears'. This could be understood as the smaller pulse may have accumulated enough energy so that it transforms into a taller pulse with a broader spectrum.

**Dual-comb lasing** When gradually decreasing the pump power from the above states, the CW peak on top of the optical spectrum could be removed by carefully decreasing the pump power. Due to the well-known hysteresis effect, the laser would remain mode-locked at a lower power than its starting threshold. The laser can be adjusted into a state emitting just one 'fs' pulse and one 'ps' one. Additional measurements and analyses are carried out to characterize this dual-pulse output. Its optical spectrum at a pump power of 14.7 mW is shown in Fig. 4(a). Despite of its special shape, it's shown that it can be fitted with two standard solitons' spectra of different bandwidths. The sum of a 4.4-nm-bandwidth spectrum (blue dashed line) and a 1.3 nm one (red dashed line) separated by 0.8 nm close matches the measured data in Fig. 4(a). The inset shows an excellent agreement between the fitted narrow-bandwidth spectrum and that at a pump power of 9.2 mW. Intensity autocorrelation measurement is done by sending both pulses into an autocorrelator. The measured autocorrelation trace matchs the sum of two individual autocorrelation functions of two $sech^2$ pulses with 570 fs and 2.0 ps pulsewidth, which are estimated based on the fitted optical spectra respectively in Fig. 4(a) and assuming transformed-limited solitons. In this case, the 'fs' pulse's peak power is estimated to be one order of magnitude larger than that of the 'ps' pulse. Obviously, the peak of autocorrelation trace is mostly dominated by the 'fs' pulse and the broader tails illustrates the presence of the longer and weaker 'ps' pulse as observed in Fig. 3(b). From the RF spectrum shown in Fig. 4(c), the existence of two repetition rates closely spaced to each other with the difference $\Delta f$ of 45 Hz further proves the simultaneous stable lasing of one femtosecond and one picosecond pulses in the cavity. The value of $\Delta f$ is in good agreement with the cavity dispersion. The smaller peaks are beat notes generated by the photodetector. Therefore, these observations confirm that two pulse trains with overlapped spectra and pulsewidths different by a factor of more than 3 are simultaneously generated.

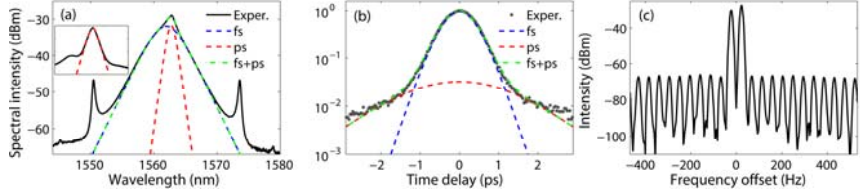

Fig. 4. (a) Measured and $sech^2$-fitted optical spectra; (b) autocorrelation trace and calculated ones based on the spectral information; (c) RF frequency spectrum of the laser output.

By slightly tuning the state of polarization, it is observed that the center wavelength of the 'ps' pulse can be adjusted, as the maxima of spectral interference shift. Also, when stronger polarization-dependent loss is introduced, the fs pulse's spectrum under hybrid mechanism becomes wider and further shifts to shorter wavelengths due to the changed gain tilt. This may further increase the center wavelength separation and in turn result in a larger frequency difference. Accordingly, the repetition rate difference between the two pulses is found to be tunable over a range of about 30 to 80 Hz. Fig. 5 shows a case where the spectra of the two pulses are 5.2 and 0.8 nm-bandwidth, respectively. A center wavelength separation of 1.4 nm leads to a $\Delta f$ of 78 Hz.

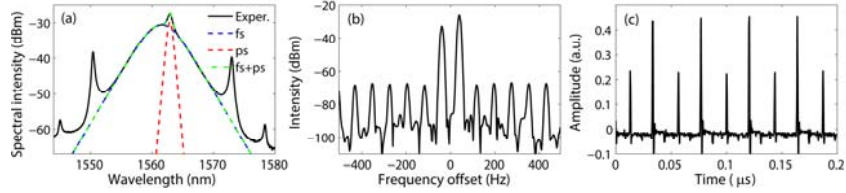

Fig. 5. (a) Experimental and $sech^2$ fitted optical spectra; (b) RF frequency spectrum and (c) waveform of the laser output when $\Delta f$ = 78 Hz.

**Asynchronous coherent sampling and interferogram** Two coherent pulse trains with a stable $\Delta f$ could asynchronously sample each other. To further verify the characteristics of the dual-comb output, it is split into two paths and then recombined by using 2*2 50:50 optical couplers (OCs) (see Fig. 6(a)). The two outputs of the coupler are sent to a balanced photodetector (BPD, Thorlabs PDB420C), for the background-free interferogram measurement. An NI-5122, 14-bit ADC at a sampling rate of 100 MS/s is used to digitize the signal. At $f_{rep}$ = 22.975 MHz and $\Delta f$ = 63 Hz, the following measurements are made with a corresponding sampling step of 119 fs. The output from BPD shown in Fig. 6(b) is similar to the electric field cross-correlation scaled by a factor of $f_{rep}/\Delta f$. Its envelope can be fitted with a 1.08 μs-wide $sech^2$ shape, corresponding to a temporal width of 2.96 ps. Optical spectral shape in Fig. 6(c) is retrieved by taking a Fourier transform of the temporal interferogram as in the dual-comb spectroscopy measurement. It is obtained by averaging over 34 interferograms. The signal quality indicates good stability between comb lines of the dual-comb output. Its shape matches the calculated product of the dual-comb spectra, mostly determined by the 'ps' pulse, which is 1.2 nm (i.e. 150 GHz) wide in this measurement.

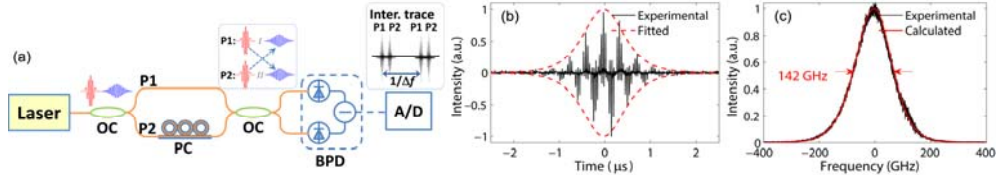

Fig. 6. (a) Experimental setup for asynchronous sampling; (b) measured interferogram trace and (c) its Fourier transform.

## 4. Conclusion

We demonstrate a unidirectional dual-comb laser that can asynchronously emit multiple ultrashort pulses with quite different pulsewidths. It is believed that the additional linear and nonlinear pulse shaping mechanisms introduced into the cavity lead to this unusual bifurcation of pulse states at elevated pump powers. Good spectral coherence due to the completely common-path cavity is demonstrated, which is evident from the asynchronous sampling and corresponding dual-comb spectroscopy result.

## Acknowledgements

This work at Beihang University was supported by 973 Program (2012CB315601) and NSFC (61435002/61521091).